\documentclass[ba]{imsart}
\pubyear{2025}
\volume{TBA}
\issue{TBA}
\firstpage{1}
\lastpage{1}

\usepackage{tikz}
\usepackage{amsmath}
\usepackage{amssymb}
\usepackage{amsthm}
\usepackage{mathtools}
\usepackage{natbib}
\usepackage[dvipsnames]{xcolor}
\usepackage[colorlinks,citecolor=blue,urlcolor=blue,filecolor=blue,backref=page]{hyperref}
\usepackage{graphicx}

\usepackage{subcaption}
\usepackage{caption}

\usepackage{fancyhdr}

\startlocaldefs
\endlocaldefs

\begin{document}


\begin{frontmatter}
\title{Integrating Expert Knowledge and Recursive Bayesian Inference: A Framework for Spatial and Spatio-Temporal Data Challenges}
\runtitle{}

\begin{aug}
\author{\fnms{Mario} \snm{Figueira}\thanksref{addr1,t1}\ead[label=e1]{Mario.Figueira@uv.es}},
\author{\fnms{David} \snm{Conesa}\thanksref{addr1}\ead[label=e2]{}},
\author{\fnms{Antonio} \snm{López-Quílez}\thanksref{addr1}\ead[label=e3]{}} \and
\author{\fnms{H\r{a}vard} \snm{Rue}\thanksref{addr2}\ead[label=e4]{}}

\runauthor{}

\address[addr1]{Department of Statistics and Operations Research, Universitat de València. C/ Dr. Moliner, 50 46100 Burjassot (València), Spain.}

\address[addr2]{CEMSE Division, King Abdullah University of Science and Technology, Kingdom of Saudi Arabia.}

\thankstext{t1}{\printead{e1}}

\end{aug}

\begin{abstract}
{

Integrating heterogeneous data sources and expert knowledge is essential for overcoming data scarcity and enhancing estimation accuracy. Two main frameworks naturally arise to perform the integration of these multiple sources: sequential Bayesian inference and integrated models. The first one consists of updating posterior information in a sequential data analysis procedure, without the need to reanalyze previous data when new data become available. The second one consists of bringing together diverse sources of information in a joint inferential analysis through hierarchical Bayesian models. Within the context of the first framework, we propose a recursive inference method grounded in the methodological principles of INLA, designed to handle spatial and spatio-temporal problems, although its applicability is not limited to these cases, as the procedure is general in nature. Within the integrated models framework, we also present a comprehensive approach to address change of support issues that arise when combining heterogeneous information sources, developing a typology that classifies such changes as spatial, temporal, spatio-temporal, or categorical. Both frameworks can be combined, as there is neither a theoretical nor a practical incompatibility preventing their joint use. Finally, detailed examples are provided to illustrate clear and replicable procedures for combining heterogeneous data sources with change of support and recursive inference.
}

\end{abstract}


\end{frontmatter}

\section{Introduction}\label{Introduction}

{ Expert knowledge and external sources of information are essential for addressing data scarcity, improving estimations by integrating diverse information sources and enhancing model robustness in contexts where data is limited. They play a crucial role in combining heterogeneous datasets, reconciling inconsistencies, and guiding inference when empirical data is incomplete or uncertain.  In this regard, expert knowledge and external sources (e.g. computer or deterministic models) are a key approach to formalizing heterogeneous information into comprehensive statistical models. It has proven essential across multiple applied domains---such as ecology, environmental sciences, public policy, and risk management---where direct observations may be limited or unreliable \citep{Kynn_Elicitation_2008, Kunhert_ExpertKnowledgeEcology_2010, Morgan_2014_ElicitationPolicy, Zhang_ExpertElicitationShippingAccidents_2016}. Moreover, these external sources effectively complement ``big data'' analytics by filling informational gaps and strengthening weakly informed model components. Such data integration is particularly valuable in integrated \citep{Ovaskainen_2011_JointSpeciesModels, TaylorRodriguez_2017_JointSpeciesDistribution, Paradinas_CombiningFisheryDataISDM_2023} and fusion models \citep{Wang_GeneralizedFusionModelPointArea_2018, Moraga_2024_ChangeOfSupport}, where diverse data sources help reconcile conflicting or sparse data information. Ultimately, combining empirical data with expert judgment---through expert knowledge and computer models \citep{OHagan_BayesianCalibrationComputerModels_2001, OHagan_UncertainJudgements_2006, Higdon_ComputerModelCalibrartion_2008, Morgan_2014_ElicitationPolicy, Zhang_ExpertElicitationShippingAccidents_2016, Wang_GeneralizedFusionModelPointArea_2018, Cadham_2021_ExpertElicitation}---provides a principled way to manage uncertainty and improve inference in complex systems.}

{ The origin of expert knowledge and external information sources can be diverse, encompassing expert elicitation procedures, computational or deterministic models, citizen science initiatives, and auxiliary datasets obtained from other studies or institutions. Such external information may also originate from models developed in previous experiments or data analyzes related to the current study, where the estimation of certain effects is assumed to be comparable to those in the target hierarchical Bayesian model. When integrating these different sources of information into a statistical framework, two general strategies naturally emerge: sequential inference and integrated modeling. The former is based on the use of a sequential inference strategy, which in turn leverages inferential results from previous analyzes or experiments to obtain estimates that can be incorporated as prior information in the target model. In particular, posterior distributions derived from these preceding models can be used to inform the prior distributions of parameters and hyperparameters in the main model. The second strategy (integrated modeling), alternative data sources---such as citizen science records, institutional databases, or outputs from computational and deterministic models---are used to fit auxiliary models that share components with the target Bayesian hierarchical model, thereby improving parameter estimation and enhancing model robustness.}

{ In this context, expert knowledge or external information may be provided at the same scale as the current data, or at a different scale. This is particularly relevant when dealing with spatial and spatio-temporal models, where the data may show a change in support or misalignment \citep[see, for example, the works by][]{Gelfand_2001_ChangeOfSupport, Gobbi_ExpertBasedConservation_2012, Kaurila_SDMExperElicitationBayesCalibration_2022, Alahmadi_2025_MisalignedHelath}. The latter deals with a change of support between the health data, presented at an aggregated area scale, and environmental variables, presented at a point-reference scale, proposing an up-scaling model for the environmental pollutant information to the area scale of the health data.}

{ In line with all this, we present a comprehensive approach to perform integration of expert knowledge and external information (i.e.,  integrating heterogeneous information sources) using sequential inference strategies and/or integrated models. Our aim is twofold. Firstly, we present how to perform recursive Bayesian inference within the context of the Integrated Nested Laplace Approximation \citep[INLA,][]{Rue_INLABaseArticle_2009}. This recursive approach allows for the fitting of models sequentially as new data sources become available while properly accounting for the estimation of both parameters and hyperparameters. This is advantageous due to the sequential nature of Bayes' theorem because it allows new information to be incorporated into the inference sequence without the need for complete model recomputation, as in online updating \citep{Schifano_OnlineUpdating_2016} and Bayesian filtering \citep{Sarkka_2013_BayesianFiltering}. Secondly, we also present a framework that explicitly addresses the challenge of change of support that may arise when using integrated models to combine heterogeneous data sources. We present this framework within the context of spatio-temporal modeling, where this issue is of special importance. In particular, when integrating heterogeneous data sources defined over different spatial (or temporal) scales. Indeed, we introduce a typology that characterizes diverse change of support scenarios involving such heterogeneous sources. The proposed framework enhances model estimation within a joint modeling structure that allows for the simultaneous integration of expert knowledge calibration and uncertainty analysis alongside observational data under change of support conditions. Importantly, the calibration and uncertainty analysis of expert knowledge are not limited to subjective expert judgments but can also incorporate information from simulators or computer models---deterministic models that emulate the underlying physical or ecological processes and can be seamlessly integrated into the framework as expert-provided data.
}

{
In summary, Section 2 provides a brief overview of the general framework and the different approaches for integrating multiple sources of information, either through integrated models or sequential inference. Section 3 reviews previous work on sequential inference and introduces a new approach for implementing recursive Bayesian inference within the INLA framework. Section 4 describes how to integrate heterogeneous sources of information through joint models, discussing the various types of integration and the methodological considerations involved, including a detailed explanation of how to handle change of support. Section 5 presents concrete examples that illustrate the proposed procedures in a clear, concise, and reproducible manner. Finally, the conclusions discuss the scope of the proposed methods and the potential limitations encountered in their implementation.
}

\section{Combining heterogeneous information sources}
\label{sec:inference_expert_integration_framework}

{ The integration of different data sources involves combining information originating from experimental designs, scientific surveys, citizen science initiatives, and institutional databases, among others. Additionally, such information may derive from expert knowledge contributed by technicians, policymakers, or other individuals recognized as experts, obtained through structured elicitation procedures. In this sense, as above mentioned, two general strategies naturally emerge for the integration of these heterogeneous sources of information.}

{ The first strategy relies on sequential (or recursive) updating of the different sources of information \citep{Hooten_RecursiveBayesianInference_2021}. However, this information is not always available at the same time as the data being analyzed, and it may become available only during subsequent surveys. In such cases, it is preferable to incorporate new information into the model for the latest data without reprocessing all previously analyzed information. This can be accomplished through different inferential approaches, such as online updating within a frequentist framework \citep{Schifano_OnlineUpdating_2016}, sequential Bayesian inference \citep{Scott_ConsensusMonteCarlo_2016, Hooten_RecursiveBayesianInference_2021, Kessler_2023_OnSequentialBayesianInference}, or continual learning methods \citep{Wang_2024_ContinualLearning}. 

In this work, we focus on the sequential framework by proposing a recursive Bayesian inference procedure. Suppose for instance that we have two datasets related to the same phenomenon, $\mathbf{y}_a$ and $\mathbf{y}_b$, and they become available at different times. A natural approach within Bayesian inference is to perform a sequential analysis of them, as they become available. This implies fitting the first dataset with a given model structure $\mathcal{M}$, and use the posterior distributions to define the prior distributions for the following dataset following the same model structure $\mathcal{M}$. The hierarchical Bayesian model $\mathcal{M}$ is characterized by the likelihood $\ell$, the link function $g$, the linear predictor $\boldsymbol\eta$ defined as $\boldsymbol\eta = \mathbf{A} \mathbf{x} = g(\mathbb{E}(\mathbf{y}))$, the latent field $\mathbf{x}$, the design matrix $\mathbf{A}$ and the hyperparameters $\boldsymbol\theta = (\boldsymbol\theta_1, \boldsymbol\theta_2)$, related respectively to the likelihood, $\boldsymbol\theta_1$, and the latent field, $\boldsymbol\theta_2$. Following this notation, this sequential scheme is synthesized in Figure \ref{fig:Example_Sharing_Observational_Observational}.}

{ The second strategy involves the use of integrated (or joint) models \citep{Ovaskainen_2011_JointSpeciesModels, Paradinas_CombiningFisheryDataISDM_2023}, in which each dataset is evaluated with its own model, while certain components are shared across them. This structure allows for improved parameter estimation and enhances the overall robustness of the inference. In this context, we may have a primary dataset that we wish to evaluate, and, given the scarcity of such data, it may be necessary to rely on auxiliary datasets related to the phenomenon of interest but for which more extensive information is available. For instance, data concerning the abundance of a given species can be complemented by presence-absence data. Furthermore, information obtained from oceanographic surveys, which could be less abundant, can be complemented with data derived from commercial catch campaigns, which could be more informative. By combining these different sources of information and sharing parameters $\mathbf{x}^*$ across models ($\mathcal{M}_a, \mathcal{M}_b$), $\mathbf{x}^*\in \mathcal{M}_a$ and $\mathbf{x}^* \in \mathcal{M}_b$, the analysis operates within the framework of integrated models. Each of the shared parameters $\mathbf{x}^*$ can be scaled up to an $\alpha_j \in \boldsymbol\alpha$ parameter to be estimated. In Figure~\ref{fig:Example_Sharing_Observational_AuxiliarModels} is shown the scheme combining two observational complementary datasets within the integrated modeling approach.}

{
\linespread{1.}
\begin{figure}
    \centering
    \begin{tikzpicture}
    \node at (-1,-1.5) [circle,draw] (1) {$\mathbf{y}_{a}$};
    \node at (7.1,-1.5) [rectangle,draw] (2) {$\everymath={\displaystyle}
    \begin{array}{c}
    \mathbf{x} \sim \pi(\mathbf{x}) \\[0.2cm]
    \boldsymbol\theta \sim \pi(\boldsymbol\theta) \\
    \end{array}$};
    
    \node at  (3,-1.5) [rectangle,draw] (3) {$\everymath={\displaystyle}
    \begin{array}{c}
    \mathbf{Y}_a \mid \boldsymbol\eta, \boldsymbol\theta_1 \sim  \ell(\mathbf{y}_{a} \mid \boldsymbol\eta_a,\boldsymbol\theta_1)\\[0.2cm]
    g\left[\mathbb{E}(\mathbf{Y}_a)\right]  =  \boldsymbol\eta_a(\mathbf{A}_a, \mathbf{x}, \boldsymbol\theta_2)\\[0.2cm]
    \mathbf{x} \sim \pi(\mathbf{x} \mid \boldsymbol\theta_2)\ \\[0.2cm]
    \boldsymbol\theta \sim \pi(\boldsymbol\theta_1, \boldsymbol\theta_2)\ \\
    \end{array}$};
    
    \node at (2.95,.35) (A) {\textbf{Observational Model}};
    
    \draw[-to] (1) to [out=0,in=180,looseness=0.5] (3);
    \draw[double, -to] (2) to [out=180,in=0] (3);
    
    \node at (-1,-6.2) [circle,draw] (4) {$\mathbf{y}_{b}$};
    \node at (7.7,-6.2) [rectangle,draw] (5) {$\everymath={\displaystyle}
    \begin{array}{c}
    \pi(\mathbf{x} \mid \mathbf{y}_{b}) \\[0.2cm]
    \pi(\boldsymbol\theta \mid \mathbf{y}_{b})\\
    \end{array}$};
    
    \node at  (3,-6.2) [rectangle,draw] (6) {$\everymath={\displaystyle}
    \begin{array}{c}
    \mathbf{Y}_b \mid \boldsymbol\eta_b, \boldsymbol\theta_1 \sim  \ell(\mathbf{y}_{b}|\boldsymbol\eta_b,\boldsymbol\theta_1) \\[0.2cm]
    g\left[\mathbb{E}(\mathbf{Y}_b)\right] = \boldsymbol\eta_b(\mathbf{A}_b, \mathbf{x}|\boldsymbol\theta_2) \\[0.2cm]
    \mathbf{x} \sim \pi(\mathbf{x}_b, \mathbf{x}^* \mid \boldsymbol\theta_2)\ \\[0.2cm]
    \boldsymbol\theta \sim \pi(\boldsymbol\theta_1, \boldsymbol\theta_2) \\
    \end{array}$};
    
    \node at (3.1,-8.3) (A) {\textbf{Previous Observational Model}};
    
    \draw[-to] (4) to [out=0,in=180,looseness=0.5] (6);
    \draw[double, -to] (6) to [out=0,in=180] (5);
    
    \node at (7.7,-3.8) [circle,draw] (Up) {$\begin{array}{c}
        \mathbf{Prior} \\
        \mathbf{setting} \\ 
        \mathbf{process} 
    \end{array}$};
    
    \draw[dotted,-to] (Up) to [out=45,in=0,looseness=1.5] (2);
    \draw[dotted,-to] (5) to [out=90,in=-90,looseness=0.5] (Up);
    
    \end{tikzpicture}
    \caption{ Implementation scheme of the sequential updating strategy for integrating different sources of expert or external information.}
    \label{fig:Example_Sharing_Observational_Observational}
\end{figure}
}

{
\linespread{1.}
\begin{figure}
    \centering
    \begin{tikzpicture}
    \node at (-1,-1.5) [circle,draw] (1) {$\mathbf{y}_{a}$};
    
    \node at  (3,-1.5) [rectangle,draw] (3) {$\everymath={\displaystyle}
    \begin{array}{c}
    \mathbf{Y}_a \mid \boldsymbol\eta_b, \boldsymbol\theta_a, \boldsymbol\alpha \sim  \ell(\mathbf{y}_{a}|\boldsymbol\eta_a,\boldsymbol\theta_a) \\[0.2cm]
    g\left[\mathbb{E}(\mathbf{Y}_a)\right]  =  \boldsymbol \eta_a(\mathbf{A}_a, \mathbf{x}^*_a, \boldsymbol\alpha, \boldsymbol\theta_{a2}) \\[0.2cm]
    \mathbf{x}^*_a \sim \pi(\mathbf{x}_a, \mathbf{x}^*|\boldsymbol\theta_{a2},\boldsymbol\alpha)\ \\[0.2cm]
    \boldsymbol \theta_a \sim \pi(\boldsymbol\theta_{a1}, \boldsymbol\theta_{a2}, \boldsymbol\alpha)\ \\
    \end{array}$};

    \node at (2.95,.35) (A) {\textbf{Observational Model} ($\mathcal{M}_a$)};
    
    \draw[-to] (1) to [out=0,in=180,looseness=0.5] (3);

    \node at (-1,-5.8) [circle,draw] (4) {$\mathbf{y}_{b}$};

    \node at  (3,-5.8) [rectangle,draw] (6) {$\everymath={\displaystyle}
    \begin{array}{c}
    \mathbf{Y}_b \mid \boldsymbol\eta_b, \boldsymbol\theta_b \sim  \ell(\mathbf{y}_{b}|\boldsymbol\eta_b,\boldsymbol\theta_b) \\[0.2cm]
    g\left[\mathbb{E}(\mathbf{Y}_b)\right]  =  \boldsymbol\eta_b(\mathbf{A}_b, \mathbf{x}^*_b, \boldsymbol\theta_2) \\[0.2cm]
    \mathbf{x}^*_b \sim \pi(\mathbf{x}_b, \mathbf{x}^*|\boldsymbol\theta_b)\ \\[0.2cm]
    \theta_b \sim \pi(\boldsymbol\theta_{b1}, \boldsymbol\theta_{b2})\ \\
    \end{array}$};

    \node at (3.1,-7.9) (A) {\textbf{Observational Model} ($\mathcal{M}_b$)};

    \draw[-to] (4) to [out=0,in=180,looseness=0.5] (6);
    
    \node at (3,-3.5) [rectangle,draw] (C) {\textbf{Linked through  $\eta(\mathbf{x},\boldsymbol\theta)$}};
    \draw[-] (3) to [out=270,in=90] (C);
    \draw[-] (C) to [out=270,in=90] (6);

    \end{tikzpicture}
    \caption{ Implementation scheme of integrated approach strategy for integrating different sources of auxiliar or complementary information.}
    \label{fig:Example_Sharing_Observational_AuxiliarModels}
\end{figure}
}

{ The two main contributions of this work for dealing with the integration of heterogeneous data are as follows. First, we propose a recursive inference framework grounded in the methodological principles of INLA, designed to handle spatial and spatio-temporal problems, although its applicability is not limited to these cases, as the procedure is general in nature. Second, we provide a comprehensive approach to address change of support issues that arise when combining heterogeneous information sources, developing a typology that classifies such changes as spatial, temporal, spatio-temporal, or categorical. It is important to note that, although the proposed approach to change of support is presented within the context of integrated models for simplicity, there is no conceptual or methodological constraint preventing the use of integrated models within recursive inferential processes. Consequently, the proposed framework for addressing change of support is fully compatible and jointly implementable with the recursive inference proposal.}

{ As above mentioned, the approach proposed in this paper for integrating expert or external knowledge and the implementation of recursive inference to allow an updatable procedure of the model, is grounded within the framework of the Integrated Nested Laplace Approximation \citep{Rue_INLABaseArticle_2009, VanNiekerk_2023_NewINLA} in the \texttt{R-INLA} software \citep{Martins_2013_RINLA}. INLA is a deterministic approximation approach deeply rooted in Gaussian Markov Random Field (GMRF) theory \citep{Rue_GMRF_2005} for Bayesian inference \citep{Rue_INLABaseArticle_2009, VanNiekerk_2023_NewINLA} and Latent Gaussian Models (LGM). INLA focuses on estimating the marginal posterior distributions of the model parameters, allowing the computation of marginal likelihoods and standard goodness-of-fit metrics such as Deviance Information Criterion (DIC) \citep{Spiegelhalter_DIC_2002}, Watanabe-Akaike Information Criterion (WAIC) \citep{Watanabe_WAIC_2013}, Conditional Predictive Ordinates (CPO) \citep{Pettit_CPO_1990}, and leave-group-out cross validation \citep{Liu_2025_LGOCV}, all of which are essential for model evaluation and comparison. Although we focus and ground our method in the INLA underpinnings, other sequential inferential procedures have been proposed based on MCMC methods \citep{Scott_ConsensusDistributedComputation_2017, Rendell_GlobalConsensus_2021} or in the properties of samplers for the MCMC, taking advantage of features from the Metropolis-Hastings sampler \citep{Lunn_2013_FullyBayesianTwoStages, Hooten_RecursiveBayesianInference_2021}, the Weierstrass sampler \citep{Wang_2014_ParallelMCMCWeierstrass}, expectation propagation \cite{Guhaniyogi_2022_DistributedBayesianInference} and others.}

\section{Sequential and recursive inference}

{ In this section, we introduce different strategies for integrating different sources of information within Bayesian inference, with a particular focus on implementation using the INLA approach. These strategies can be applied using the INLA methodology through the \texttt{R-INLA} software \citep{Martins_2013_RINLA}, combined with other methods such as Monte Carlo \citep{Berild_ImportanceSamplingINLA_2022} and MCMC \citep{GomezRubio_MCMCwithINLA_2018}, or implemented entirely using MCMC approaches.}

{ The section is divided into two parts. The first part discusses previous proposals for updating prior information. In particular one of them, when combined with a consensus approach \citep{Scott_ConsensusMonteCarlo_2016}, it enables sequential consensus inference \citep{Figueira_2025_SequentialConsensusInference}. The second part introduces our proposed approach for performing recursive inference within the INLA framework, without relying on the assumptions and approximations of the previously discussed sequential consensus approach. More importantly, it can be applied to any model that can be implemented within the INLA framework, provided that the successive stages of the recursive procedure satisfy the standard conditions required for any such model.} 

\subsection{Prior updating and sequential consensus}

In a general sense, { within a sequential procedure prior updating refers} to the process of refining a prior distribution based on new knowledge or relevant information. This is a fundamental concept in Bayesian statistics, as Bayes' theorem itself is built upon the principle of updating prior beliefs to obtain a posterior distribution. In this subsection, we present different strategies for updating model information: direct prior updating (\textit{prior updating by moments} and \textit{full prior updating}) as well as an algorithm that integrates sequential prior updating with consensus to update the entire model's information \citep{Figueira_2025_SequentialConsensusInference}.

\subsubsection{Prior updating by moments}

{ This strategy involves} updating the moments that define the kernel of the prior distribution, such as the mean and variance in a Gaussian kernel. This approach can be easily implemented in both MCMC and INLA for univariate distributions. Since INLA relies on latent Gaussian models (LGM) and the fixed parameters (the $\boldsymbol\beta$ coefficients) are expected to follow a Gaussian (or nearly-Gaussian) distribution, we can update their Gaussian prior by modifying its characteristic moments, as discussed in \cite{Figueira_2024_BayesianFeedback}. For hyperparameters in INLA, we can either update the existing kernel of the prior distribution or introduce a new one, refining it to better align with the posterior distribution from a previous model. However, for the linear coefficients $\boldsymbol\beta$, we remain constrained to Gaussian priors in the INLA framework.

This can be achieved by determining the parameters $\boldsymbol\xi$ of $\pi(x \mid \boldsymbol\xi)$ based on the moments (e.g., mean and variance) of the marginal posterior distribution $\pi(x \mid \mathbf{y})$. Alternatively, $\boldsymbol\xi$ can be estimated by minimizing the Kullback-Leibler divergence \cite[KLD,][]{Kullback_InformationTheory_1997} between the two distributions, i.e., solving $\min_{\boldsymbol\xi} \text{KLD}(\pi(x\mid \boldsymbol\xi) \lVert \pi(x\mid \mathbf{y}))$.

\subsubsection{Full prior updating}

Full prior updating enables the definition of arbitrary continuous distributions, including joint prior distributions for both parameters and hyperparameters. The goal is to define joint prior distributions $\pi(\mathbf{x} \mid \boldsymbol\xi)$ based on the posterior distributions $\pi(\mathbf{x} \mid \mathbf{y})$. Within the \texttt{R-INLA} framework, three key features facilitate hyperparameter updating: the \texttt{table} and \texttt{expression} functionalities \citep{GomezRubio_BayesianInferenceINLA_2020}, along with the \texttt{rprior} function, which streamlines the creation of custom priors. This functionality allows us to define a distribution by specifying a set of values for the random variable along with their corresponding density values, $\{x_i, \pi(x_i)\}$. It is important to note that these density values do not always directly correspond to the parameter itself but may instead relate to a transformation of the parameter. Therefore, careful consideration of the appropriate transformation, accounting for internal parameterization, is essential. However, constructing joint priors and assigning non-Gaussian priors to latent field parameters presents certain challenges.

To address these challenges, integrating MCMC methods with INLA provides a viable solution \citep{GomezRubio_MCMCwithINLA_2018}. For instance, a Metropolis-Hastings algorithm can be implemented to jointly update a set $\boldsymbol\nu$ of latent field parameters ($\mathbf{x}$) or hyperparameters ($\boldsymbol\theta$), where $\boldsymbol\xi = \{\mathbf{x}, \boldsymbol\theta\}$. This is achieved by defining a suitable proposal distribution $q(\cdot \lvert \cdot)$ and computing the acceptance probability:
$$
\alpha = \min\left\lbrace1,\frac{\pi(\mathbf{y} \mid \boldsymbol\xi^*)\pi(\boldsymbol\xi^*)q(\boldsymbol\xi^{(j)} \mid \boldsymbol\xi^*)}{\pi(\mathbf{y} \mid \boldsymbol\xi^{(j)})\pi(\boldsymbol\xi^{(j)})q(\boldsymbol\xi^* \mid \boldsymbol\xi^{(j)})}\right\rbrace,
$$
where $\pi(\mathbf{y} \mid \boldsymbol\xi^{(j)})$ and $\pi(\mathbf{y} \mid \boldsymbol\xi^*)$ are the conditional marginal likelihoods for $\boldsymbol\xi^{(j)}$ and $\boldsymbol\xi^*$, respectively. Alternatively, Importance Sampling or Adaptive Multiple Importance Sampling \citep{Berild_ImportanceSamplingINLA_2022} can be used to perform full prior updating.

\subsubsection{Sequential Consensus inference}

The combination of prior updating procedures, particularly prior updating by moments, along with a consensus approach, allows us to define sequential algorithms for conducting Bayesian inference with INLA. The sequential consensus inference approach \citep{Figueira_2025_SequentialConsensusInference} focuses on sequentially updating the marginal distributions of the fixed effects of the latent field $\boldsymbol\beta$ and hyperparameters $\boldsymbol\theta$ according to prior updating by moments, and reaching a consensus on the latent field related to the random effects, either marginal or multivariate. Specifically, the following approximation is proposed for the marginal posterior distribution of the fixed effects:
\begin{equation}
\pi(\beta_i \mid \mathbf{y}_2, \mathbf{y}_1) \approx \pi(\mathbf{y_2} \mid \beta_i) \pi(\beta_i \mid \mathbf{y}_1),
\end{equation} 
and a similar approximation is proposed for the hyperparameters. Once the sequence is completed, a marginal consensus of the random effects is also proposed, either by the weighted sum of the $n$ parts of the sequence for the random effects of the latent field:
\begin{equation}
x_i \approx \sum_{j=1}^n w_{ij} \cdot x_{ij}, 
\end{equation}
where $w_{ij} = \tau_{ij}/\sum_{i=1}^n\tau_{ij}$ denotes the optimal weights, { and $\tau_{ij}$ represents the marginal precision of the $i$-th node of the latent field in partition $j$}. Alternatively, a multivariate consensus of the nodes of the latent field can be employed:
\begin{equation}
\pi(\mathbf{x}\mid \mathbf{y}_1, \mathbf{y}_2) \approx \pi(\mathbf{x}\mid \mathbf{y}_1) \cdot \pi(\mathbf{x}\mid \mathbf{y}_2),
\end{equation}
where both $\pi(\mathbf{x}\mid \mathbf{y}_1)$ and $\pi(\mathbf{x} \mid \mathbf{y}_2)$ are multivariate Gaussian distributions. 

The precision and mean of the consensus distribution of the posteriors are:
\begin{equation}
\begin{array}{rcl}
\mathbf{Q} & = & \mathbf{Q}_1 + \mathbf{Q}_2, \\
\boldsymbol\mu & = & (\mathbf{Q}_1 + \mathbf{Q}_2)^{-1}(\mathbf{Q}_1 \cdot \boldsymbol\mu_1 + \mathbf{Q}_2 \cdot \boldsymbol\mu_2).
\end{array}
\end{equation}
The precision matrices $\mathbf{Q}$ and vector means $\boldsymbol\mu$ for performing the multivariate consensus would be those obtained in the modal configuration for the hyperparameters at the different steps of the sequence. For the marginal consensus, the random effects are distributed according to the marginal distributions of the nodes of the latent field related to those random effects.

\subsection{Recursive Bayesian inference}

{ As mentioned above, in this subsection we introduce a new approach for performing Bayesian recursive inference within the INLA framework, without relying on the assumptions and approximations underlying the previously discussed Sequential Consensus approach \citep{Figueira_2025_SequentialConsensusInference}. More importantly, the proposed method addresses several issues identified in previous approaches and provides a foundation consistent with the assumptions of the INLA methodology. Since our proposal is grounded in the INLA framework, we begin this subsection by briefly introducing some concepts and notation required to properly explain the recursive approach.
}

INLA focuses in the posterior marginal distributions for the nodes of the latent field ($\mathbf{x}$) and for the hyperparameters ($\boldsymbol\theta = \{\boldsymbol\theta_1,\boldsymbol\theta_2\}$). However, to achieve this it needs to approximate the posterior distribution of the hyperparameters and the posterior distribution of the latent field given a set of support points $\{\boldsymbol\theta^k\}^K_{k=1}$ in the hyperparameter space \citep{Rue_INLABaseArticle_2009}. The approximation used to compute the posterior of the hyperparameters is $\tilde{\pi}(\boldsymbol\theta\mid\mathbf{y})$, defined as:
\begin{equation}
	\tilde{\pi}(\boldsymbol\theta\mid\mathbf{y}) = \left.\frac{\pi(\mathbf{y},\mathbf{x},\boldsymbol\theta)}{\pi_G(\mathbf{x}\mid \mathbf{y},\boldsymbol\theta)}\right|_{\mathbf{x=\mathbf{x}^*(\boldsymbol\theta)}},
\end{equation}
where $\tilde{\pi}_G(\mathbf{x}\mid \mathbf{y},\boldsymbol\theta)$ is the Gaussian approximation for the posterior of the latent field, and the whole expression is analyzed in the modal configuration of the posterior of the latent field under the Gaussian approximation.

To calculate the posterior distribution of the latent field, a Gaussian approximation is used, evaluated at the support points, and a low-rank correction is applied to the joint mean of the posterior distribution of the latent field under the Gaussian approximation \citep{VanNiekerk_2023_NewINLA, VanNiekerk_2024_LowRankVariational}. This correction can also be used to correct the marginal variance. The Gaussian approximation to the true $\pi(\mathbf{x} \mid \mathbf{y}, \boldsymbol{\theta})$ is done by matching the mode and the curvature around the mode of $\pi(\mathbf{x} \mid \mathbf{y}, \boldsymbol{\theta})$ using a second-order Taylor series expansion around the mode of $\boldsymbol{\mu}(\boldsymbol{\theta})$ \citep{Rue_INLABaseArticle_2009, VanNiekerk_2023_NewINLA}:
\begin{equation}
\begin{array}{rcl}
\pi_{G}(\mathbf{x}\mid \mathbf{y}, \boldsymbol\theta) & \approx & \exp(\log\pi(\boldsymbol\mu(\boldsymbol\theta)\mid \mathbf{y},\boldsymbol\theta) + \nabla \left.\log\pi(\mathbf{x}\mid \mathbf{y}, \boldsymbol\theta)\right|_{\mathbf{x}=\boldsymbol\mu(\boldsymbol\theta)} \mathbf{x} \\[1.5mm]
 & & + \frac{1}{2}\mathbf{x}^T \nabla^2\left.\log\pi(\mathbf{x}\mid \mathbf{y},\boldsymbol\theta)\right|_{\mathbf{x}=\boldsymbol\mu(\boldsymbol\theta)} \mathbf{x}.
\end{array}
\end{equation}
This posterior approximation is obtained for each support point ${\boldsymbol{\theta}^k}$, and the mean is corrected through an iterative procedure of low-rank variational Bayes correction \citep{VanNiekerk_2024_LowRankVariational}. This yields a GMRF with a corrected mean for each support point ${\boldsymbol{\theta}^k}$:
\begin{equation}
\pi_G(\mathbf{x}\mid \mathbf{y}, \boldsymbol\theta^k) = \text{GMRF}(\mu^*(\boldsymbol\theta^k), \mathbf{Q}(\boldsymbol\theta^k)),
\end{equation}
where $\boldsymbol{\mu}^*(\boldsymbol{\theta})$ is the corrected mean and $\mathbf{Q}(\boldsymbol{\theta})$ is the precision matrix. The GMRF of the latent field for each ${\boldsymbol{\theta}^k}$ is one of the outputs returned by the \texttt{inla} call in the \texttt{R-INLA} package, and it will be one of the key elements for implementing a recursive inference procedure with INLA.

Finally, two distinct strategies are used to obtain the marginals of the hyperparameters and the latent field. For the hyperparameters, interpolation is performed over the values obtained at the support points $\{\boldsymbol{\theta}^k\}$ to yield a marginal distribution for each hyperparameter. Meanwhile, to obtain the marginal of the latent field, the following approximation is defined as a mixture of the marginal distributions at each support point:
\begin{equation}
\tilde{\pi}(x_i\mid \mathbf{y})\approx \sum_{k=1}^K \pi_G(x_i\mid \boldsymbol\theta^k, \mathbf{y})\tilde{\pi}(\boldsymbol\theta^k\mid \mathbf{y})\Delta_k, 
\label{eq:INLA_latent_field_marginal}
\end{equation} 
where $\pi_G(x_i \mid \boldsymbol{\theta}^k, \mathbf{y})$ is the marginal distribution given the GMRF with the corrected mean, and $\Delta_k$ are weights associated with each support point, related to their distribution.

{ Therefore, based on the principles of the INLA methodology and on the fact that INLA relies on the conditional independence of the data process given the latent field and the hyperparameters, $\pi(y_i, y_j \mid \mathbf{x},\boldsymbol\theta) = \pi(y_i \mid \mathbf{x}, \boldsymbol\theta) \cdot \pi(y_j \mid \mathbf{x}, \boldsymbol\theta)$, we can extend this framework to define a recursive inferential procedure.}

{ Let us assume that we have} a dataset that can be split---or is naturally partitioned---into $N$ disjoint subsets $\mathbf{y}=\{\mathbf{y}_1,...,\mathbf{y}_N\}$. The core structure for performing recursive inference involves analyzing the first partition, $\mathbf{y}_1$, as usual. From this partition, we obtain the support points $\{\boldsymbol\theta^k\}_{k=1}^K$, the joint posterior density of the hyperparameters evaluated at these points, $\tilde{\pi}(\boldsymbol\theta^k \mid \mathbf{y}_1)$, and the (approximated) conditional posterior GMRF of the latent field, $\tilde{\pi}_G(\mathbf{x}\mid \mathbf{y}_1, \boldsymbol\theta^k)$, for each support point. Using these outputs, we can reuse the support points $\{\boldsymbol\theta^k\}$ and the posterior GMRF of the latent field for each support point. This allows us to perform an analysis of the next subset by fixing the posterior GMRF and running parallel calls for each support point, resembling the following expression:
\begin{equation}
\begin{array}{rcl}
\pi(\mathbf{x} \mid \boldsymbol\theta^k, \mathbf{y}) & \propto & \prod_{i=1}^N \pi(\mathbf{y}_i \mid \mathbf{x}, \boldsymbol\theta^k) \pi(\mathbf{x} \mid \boldsymbol\theta^k) \pi(\boldsymbol\theta^k), \\[1.5mm]
 & = &  \prod_{i=2}^N \pi(\mathbf{y}_i \mid \mathbf{x}, \boldsymbol\theta^k) \pi(\mathbf{x} \mid \mathbf{y_1}, \boldsymbol\theta^k),
\end{array}
\end{equation}
where for the $i-th$ step we use as prior distribution for the latent Gaussian field the posterior from the previous one ($i-1$). This allow us to compute the joint posterior distribution of the latent field. 

The next step is to compute the posterior distribution for the hyperparameters throughout the sequence. In the previous first step, we have computed the support points and the density of the joint posterior distribution at these points. To update the density at the support points in subsequent steps, we can rely on the marginal likelihood of the model. Since we are fixing the computations at these support points, the marginal likelihood in steps beyond the first is conditioned on them:
\begin{equation}
\begin{array}{rcl}
\pi(\boldsymbol\theta^k \mid \mathbf{y}) & \propto & \pi(\mathbf{y} \mid \boldsymbol\theta^k) \pi(\boldsymbol\theta^k), \\
 & = & \pi(\boldsymbol\theta^k \mid \mathbf{y}_1) \prod_{i=2}^N \pi(\mathbf{y}_i \mid \boldsymbol\theta^k),
\end{array}
\end{equation}
where $\pi(\mathbf{y}_i \mid \boldsymbol\theta^k)$ is the marginal likelihood conditioned on the set of support points $\{\boldsymbol\theta^k\}_{k=1}^K$.

Therefore, the posterior distribution $\tilde{\pi}(\boldsymbol\theta^k \mid \mathbf{y})$ computed through the recursive algorithm is:
\begin{equation}
\tilde{\pi}(\boldsymbol\theta^k \mid \mathbf{y}) \propto \left[\left.\frac{\pi(\mathbf{y}_1 \mid \mathbf{x},\boldsymbol\theta^k)\pi(\mathbf{x} \mid \boldsymbol\theta^k) \pi(\boldsymbol\theta^k)}{\pi_G(\mathbf{x}\mid \mathbf{y}_1,\boldsymbol\theta^k)}\right|_{\mathbf{x=\mathbf{x}^*(\boldsymbol\theta^k)}}\right] \cdot \prod_{i=2}^N \pi(\mathbf{y}_i \mid \boldsymbol\theta^k).
\end{equation}
From this expression, the marginal likelihood can be approximated using a numerical integration scheme (e.g., central composite design, grid exploration, etc.):
\begin{equation}
\tilde{\pi}(\mathbf{y}) = \displaystyle \int \frac{\pi(\mathbf{y}\mid \mathbf{x}, \boldsymbol\theta) \pi(\mathbf{x}\mid\boldsymbol\theta)\pi(\boldsymbol\theta)}{\pi_G(\mathbf{x} \mid \mathbf{y}, \boldsymbol\theta)} d\boldsymbol\theta \approx \sum_{k = 1}^{K} \left.\frac{\pi(\mathbf{y}\mid \mathbf{x}, \boldsymbol\theta^{k}) \pi(\mathbf{x}\mid\boldsymbol\theta^{k})\pi(\boldsymbol\theta^{k})}{\pi_G(\mathbf{x} \mid \mathbf{y}, \boldsymbol\theta^{k})}\right|_{\mathbf{x}=\mathbf{x}^*(\boldsymbol\theta^{k})}\; .
\end{equation}

The final expression for the joint distribution of the hyperparameters when performing the recursive INLA procedure is:
\begin{equation}
\begin{array}{rcl}
    \tilde{\pi}(\boldsymbol\theta^k \mid \mathbf{y}) & \propto & \displaystyle \prod_{i=2}^N \left[\frac{\pi(\mathbf{y}_i \mid \mathbf{x}, \boldsymbol\theta^k)\pi(\mathbf{x}\mid \cup_{l=1}^{i-1}\mathbf{y}_l, \boldsymbol\theta^k)}{\pi^*_G(\mathbf{x} \mid \cup_{l=1}^i\mathbf{y}_l, \boldsymbol\theta^k)}\right]_{\mathbf{x}=\mathbf{x}^*_i(\boldsymbol\theta^k)} \\
     & & \displaystyle \times \left[\frac{\pi(\mathbf{y}_1 \mid \mathbf{x}, \boldsymbol\theta^k)\pi(\mathbf{x}\mid \boldsymbol\theta^k)\pi(\boldsymbol\theta^k)}{\pi_G^*(\mathbf{x} \mid \mathbf{y}_1, \boldsymbol\theta^k)}\right]_{\mathbf{x}=\mathbf{x}^*_1(\boldsymbol\theta^k)} \\
     & = & \displaystyle \prod_{i=1}^N \left[\frac{\pi(\mathbf{y}_i \mid \mathbf{x}, \boldsymbol\theta^k)\pi(\mathbf{x}\mid \cup_{l=1}^{i-1}\mathbf{y}_l, \boldsymbol\theta^k)}{\pi^*_G(\mathbf{x} \mid \cup_{l=1}^i\mathbf{y}_l, \boldsymbol\theta^k)}\right]_{\mathbf{x}=\mathbf{x}^*_i(\boldsymbol\theta^k)} \pi(\boldsymbol\theta^k)
\end{array}
.
\end{equation}

Once the joint posterior distributions of the hyperparameters and the GMRF of the latent field are obtained, with the mean rectified by the low-rank correction, the marginal distributions can be calculated. For the hyperparameters, this is done through interpolation and integration, while for the latent field, we use the same approximation employed by INLA, as described in Eq. \ref{eq:INLA_latent_field_marginal}, where we reuse the weights $\Delta_k$ obtained in the first step of the sequence.

In the proposed procedure for performing recursive inference, the first discrepancy that may arise, compared to simultaneous inference on all the data, comes from the identification of the support points $\{\boldsymbol\theta^k\}_{k=1}^K$. The main issue stems from the potential shift in the mode of $\pi(\boldsymbol\theta \mid \mathbf{y})$ throughout the sequence, leading to a possible inadequate exploration of the posterior distribution. This could be corrected by assessing the shift in the mode. However, the non-rectifiable discrepancy lies in the fact that the computed posterior distribution density is not exactly the same as that obtained through the simultaneous analysis of all the data:
\begin{equation}
\tilde{\pi}(\boldsymbol\theta^k \mid \mathbf{y}) \propto 
\left\lbrace 
\begin{array}{l} \displaystyle
\left.\frac{\pi(\mathbf{y} \mid \mathbf{x},\boldsymbol\theta^k) \pi(\mathbf{x}\mid\boldsymbol\theta^k)}{\pi_G(\mathbf{x}\mid \mathbf{y},\boldsymbol\theta)}\right|_{\mathbf{x=\mathbf{x}^*(\boldsymbol\theta^k)}} \pi(\boldsymbol\theta^k), \quad \text{(INLA)}\\ \displaystyle
\displaystyle \prod_{i=1}^N \left[\frac{\pi(\mathbf{y}_i \mid \mathbf{x}, \boldsymbol\theta^k)\pi(\mathbf{x}\mid \cup_{l=1}^{i-1}\mathbf{y}_l, \boldsymbol\theta^k)}{\pi^*_G(\mathbf{x} \mid \cup_{l=1}^i\mathbf{y}_l, \boldsymbol\theta^k)}\right]_{\mathbf{x}=\mathbf{x}^*_i(\boldsymbol\theta^k)} \pi(\boldsymbol\theta^k).
\end{array} \right.
\end{equation}
This discrepancy would transfer to the computation of the marginal distributions of the nodes of the GMRF in the latent field, being the only non-rectifiable discrepancy. However, it is estimable and, therefore, analyzable.

\section{Dealing with change of support}

{ In this Section we propose a framework for dealing with change of support when heterogeneous datasets are analyzed together but presented at different scales. We start explaining the change of support issue along with an analysis of heterogeneous sources in the absence of change of support, setting the foundation for the proposed framework in the presence of change of support. For simplicity, this framework is explained through an integrated modeling strategy, though it could also be applied using sequential or recursive inferential procedures. }

{ The distinction between heterogeneous data sources with and without a change of support lies in whether the scales at which these sources are integrated remain consistent. In many practical situations, heterogeneous data are available at different spatial, temporal, or categorical scales, leading to potential misalignment between datasets. Such discrepancies are commonly referred to as change of support or data misalignment problems, and they pose a key challenge for the joint analysis of multi-source information. For instance, in environmental and ecological applications, it is common for some datasets to represent aggregated or coarse-scale information, while others provide fine-resolution or point-level observations \citep{Yamada_ElicitingWildlifeHabitat_2003, Gobbi_ExpertBasedConservation_2012, Aizpurua_HabitSuitabilityExpertKnowledge_2015, Febbraro_ExpertBasedCorrelativeModels_2018, Fitzgerald_ExpertElicitationHabitatSpaceTime_2021, Kaurila_SDMExperElicitationBayesCalibration_2022}. These scale inconsistencies must therefore be explicitly addressed to ensure coherent integration and accurate inference within a unified modeling framework. Consequently, one source may be available at a lower (or higher) resolution than others, requiring an integration approach that accounts for the specific type of change of support. The classification below provides a framework for assessing and integrating multi-misaligned sources: (i) spatial change of support, (ii) temporal change of support, (iii) spatio-temporal change of support, and (iv) categorical change of support. }

The first three cases can be effectively integrated through a downscaling process using a Gaussian field with a Matérn structure \citep{Moraga_GeostatisticalPointFusionModel_2017, Wang_GeneralizedFusionModelPointArea_2018}. This can be applied in one dimension for temporal downscaling, in two dimensions for spatial downscaling, or as a combination of both for spatio-temporal downscaling. In the case of categorical change of support, where categorical levels are grouped or intersected, external information should influence the aggregated category, either by informing the new grouping or by adjusting the parametric structure assessing the categorical level being aggregated.

{ In the following subsections, we examine the baseline scenario with no change of support, followed by the four different cases involving change of support  mentioned above, and propose strategies to address each situation.}

\subsection{No change of support}

{ In the case with no change of support, the most simple scenario arise when only two different sources are available at the same scale, allowing for a direct integration as illustrated in Figure \ref{fig:Example_Sharing_Observational_AuxiliarModels}. Therefore, an integrated model with no change of support can be defined by a likelihood function $\ell$ for each of the $\mathbf{y}_{a}$ and $\mathbf{y}_{b}$ datasets, within a given model structure $\mathcal{M} \equiv \{\ell, \boldsymbol\eta, \mathbf{A}, \mathbf{x}, \boldsymbol\theta\}$. As in Section 2, $\boldsymbol\eta$ represents the linear predictor, $\mathbf{A}$ is the design matrix of covariates, $\mathbf{x}$ denotes the latent Gaussian field, and $\boldsymbol\theta$ consists of the hyperparameters $\boldsymbol\theta = \{\boldsymbol\theta_1, \boldsymbol\theta_2\}$, where $\boldsymbol\theta_1$ corresponds to the likelihood parameters and $\boldsymbol\theta_2$ to the latent field parameters. 

This first data model $\mathcal{M}_a$ is combined with the other source defining its correspondent model $\mathcal{M}_b$. Since $\mathbf{y}_{b}$ is at the same scale as $\mathbf{y}_{a}$ data, it preserves the same scale for the latent effects. Consequently, the integrated model can be expressed as:
\begin{equation}
\begin{array}{c}
    y_{i,a} \sim \ell_a(\mu_{i,a},\boldsymbol\theta_{a1}),\\
    g_a(\mu_{i,a})=\eta_{i,a}=\mathbf{A}_{i}\boldsymbol\beta + \sum_{k=1}^Kf_k(z_{ik}),\\
    y_{i,b} \sim \ell_b(\mu_{i,b},\boldsymbol\theta_{b1}),\\
    g_b(\mu_{i,b})=\alpha\cdot\eta_{i,a} + u_i =\alpha\left[\mathbf{A}_{i}\boldsymbol\beta + \sum_{k=1}^Kf_k(z_{ik})\right] + u_i,\\
\end{array}
\label{eq:Non_cross_scale_model_example_1}
\end{equation}
where $\mathbf{x}=\{\boldsymbol\beta, \mathbf{f}\}$ is the latent field composed by the fixed $\boldsymbol\beta$ and random effects $\mathbf{f}$. In this integrated model, the linear predictor of the $\mathcal{M}_a$ model, $\boldsymbol\eta_{a}$, is incorporated into the other model and scaled by a parameter $\alpha$. It is worth noting that the incorporation of the linear predictor of $\mathcal{M}_a$ in the linear predictor of $\mathcal{M}_b$  can be performed the other way around. }

The scaling parameter $\alpha$ is expected to be close to $1$ if the effect for both models is consistent. However, $\alpha$ can also be defined as a spline function, allowing for an assessment of its deviation from $1$ across different predictor values. The term $u_i$ represents a random effect that captures the residual uncertainty in the complementary model that is not explained by the scaled predictor.

{ In the particular case when one of the datasets comes from a elicitation process of expert information, it is crucial to analyze the hyperparameter associated with the variance (or precision) of the corresponding expert model, as recommended by \cite{OHagan_UncertainJudgements_2006}. This analysis helps quantify the uncertainty introduced by the expert information. If we define the precision of the likelihood for the expert data by $\tau_{i,exp}$, the  expert  may provide an estimate of the deviation or precision for each data point, $\{y_{i,exp}, \phi_{i,exp}\}$. The corresponding uncertainty can be represented as:
$$\tau_{i,exp}=\psi\cdot\phi_{i,exp},$$
where $\phi_{i,exp}$ is the expert-provided deviation or precision for each data point $y_{i,exp}$ and $\psi$ is a parameter to be estimated (or defined by the elicited prior in case the elicitation procedure allows the expert to provide a full elicitation). This elicitation process can incorporate information from multiple expert  sources, including expert-informed models, computational models, or specialized professionals. In such cases, it is beneficial to explore potential correlations among these  experts, as highlighted by \cite{Albert_CombiningExpertElicitation_2012}. Within the INLA framework, these correlations can be conceptualized as a multivariate Gaussian error term.

Extending the previous model, if we have multiple sources of information $m\in \{1,...,M\}$, each with their own scaling parameter $\alpha_m$, the model can be expressed as:
\begin{equation}
\begin{array}{c}
    Y_{i,a} \sim \ell_a(\mu_{i,a},\boldsymbol\theta_1),\\
    g_a(\mu_{i,a}) = \eta_{i,a}=\mathbf{A}_i\boldsymbol\beta + \sum_{k=1}^Kf_k(z_{ik}),\\
    Y_{i,b1} \sim \ell_{b1}(\mu_{i,b1},\boldsymbol\theta_{b1,1}),\\
    g_{b1}(\mu_{i,b1}) = \alpha_1\cdot\eta_{i,a} + u_{i1} + u_{iB} = \alpha_1\left[\mathbf{A}_i\boldsymbol\beta + \sum_{k=1}^Kf_k(z_{ik})\right] + u_{i1} + u_{iB},\\
    \vdots \\
    Y_{i,bM} \sim \ell_{bM}(\mu_{i,bM},\boldsymbol\theta_{bM,1}),\\
    g_{bM}(\mu_{i,bM}) = \alpha_M\cdot\eta_{i,a} + u_{iM} + u_{iB} = \alpha_M\left[\mathbf{A}_i\boldsymbol\beta + \sum_{k=1}^Kf_k(z_{ik})\right] + u_{iM} + u_{iB},\\
\end{array}
\label{eq:Non_cross_scale_model_example_2}
\end{equation}
where each complementary source of information $m$ has its own scaling parameter $\alpha_m$. As above mentioned, the scaling structure can be made more flexible by defining it as a spline, allowing the scaling value to vary depending on the linear predictor. The term $u_{iB}$ captures the correlation among the complementary sources linear predictors, assuming a Multivariate Normal (MVN) distribution:
$$\mathbf{u}_{B}\sim \mathbf{MVN}(\mathbf{0},\Sigma_B),$$ 
where the diagonal elements of the covariance matrix are defined as $\Sigma_{ii}=1/\tau_i$, and the off-diagonal elements $(i\neq j)$ are $\Sigma_{ij}=\rho_{ij}/\sqrt{\tau_i\tau_j}$. It is straightforward to extend the latter formula to capture the correlation of the so called complementary sources ($\mathbf{Y}_{b1}, ...., \mathbf{Y}_{bM}$) with the $\mathbf{Y}_a$ source of information.

It is important to note that real case studies often involve complex models, where the different sources of information may follow different likelihood functions. A common example in ecological studies—such as habitat suitability or fisheries management—occurs when one source data are presence/absence-based (Bernoulli distribution), while other source is related to the probability of presence (Beta distribution). In such cases, when data distributions differ, the modeled quantity must remain compatible across likelihoods. For instance, the probability in a Bernoulli model must align with the probability in a Beta model. In some scenarios, linear predictors within the likelihood functions must be transformed to ensure that the shared information between them is correctly defined.

In what follows, we address the cases where a change of support arises between different sources of information. For simplicity, we restrict our discussion to the case of two sources, although the extension to an arbitrary number of sources is straightforward.}

\subsection{Spatial change of support}

Spatial change of support implies that the spatial structures of the different sources information do not match. This can be assessed based on whether the information is continuously structured or organized into areas (or networks).

The different scenarios that we may encounter for spatial change of support are: (i) geostatistical-areal change of support, where observations from one source of information are continuous geostatistical data, $s_i\in \mathcal{D}\subseteq \mathbb{R}^2$, and the other source of information is structured into areas, $C_i\subset \mathcal{D}$, and (ii) area-area change of support, where both sources of information are structured in areas, $B_i\in \mathbf{B}$ and $C_i \in \mathbf{C}$ respectively, but these spatial partitions are different ($\mathbf{B} \neq \mathbf{C}$).

In the case of geostatistical-areal change of support we propose to implement the development around the fusion or melding models framework \citep{Gelfand_2001_ChangeOfSupport, Berrocal_2012_SpacetimeFusion, Wang_CombiningHeterogeneousDataFusionModels_2021, Villejo_SpatiotemporalDataFusion_2023}. The core idea is to use a continuous Gaussian field $\mathbf{u}(\mathbf{s})\sim \text{GF}(\mathbf{0},\mathbf{Q})$ in the study area $\mathcal{D}$, similar to a standard geostatistical model. However, for aggregated data, the Gaussian field is integrated within each corresponding area to obtain either the mean aggregated effect or the total aggregated effect. This involves defining a set of integration points ($s_k$) to approximate the integral via numerical integration. Moreover, we can compute the mean aggregated effect as $\iint_{s\in C_i}u(s)/|C_i|ds\approx \sum_{s_k\in C_i} u(s_k)/n \; : \; s_k\in C_i$, where we consider that each integration point has associated the same area, or we can compute the total aggregation $\iint_{s\in C_i}u(s)ds$. Depending on the modeling context, other aggregation rules may also be appropriate—for example, summing the Gaussian field values when the process represents a total quantity rather than an average.

A model to deal with this change of support can be defined as in the previous section. This integrated model consists of a likelihood $\ell_a$ for the observations $\mathbf{y}_{a}$ in a continuous study region $\mathcal{D}$, and a set of observations $\mathbf{y}_{b}$ structured in a partition of the study region $\mathcal{D}= \cup_{k \in K} C_k$. For this case, the integration model we propose, assuming that both sources of information data share the same core underlying process $\eta_{i,a}$, while incorporating also the additional component $u_j$ to account for the residual uncertainty, can be expressed as follows:
\begin{equation}
\begin{array}{rcl}
    Y_{i,a} & \sim & \ell_a(\mu_{i,a},\tau_{a}),\\
    g_a(\mu_{i,a}) & = &\eta_{i,a}=\mathbf{A}_i\mathbf{x} + \mathbf{A}_{u,i}(s) \mathbf{u}(s|\rho,\sigma),\\
    Y_{j,b} & \sim & \ell_b(\mu_{j,b},\tau_{b}),\\
    g_b(\mu_{j,b}) & = &\alpha\cdot\iint_{s\in C_j} \frac{\eta_{j,a}(s)}{|C_j|}ds + u_j, \\
    & = &\alpha\left[\sum_{s_k\in C_j}\frac{\mathbf{A}(s_k)\boldsymbol\beta}{n(s_k\in C_j)} + \sum_{s_k\in C_j} \frac{\mathbf{A}_u(s_k)\mathbf{u}(s_k|\rho,\sigma)}{n(s_k\in C_j)} \right] + u_j,\\
\end{array}
\label{eq:spatial_change_support_1}
\end{equation}
where we are integrating the effect associated with the first predictor in each region where $\mathbf{y}_b$ information is available. If the values of the covariates are not available at the required integration points, sub-models can be constructed for each covariate and incorporated into the first predictor, similar to an Error model \citep{Muff_ErrorModelINLA_2015}, or in a two-stage process \citep{Villejo_SpatiotemporalDataFusion_2023}.

In the case of areal-areal change of support, we propose the same approach but perform the integration of the continuous field in both predictors. Following the structure of the previous example, we can present it as
\begin{equation}
\begin{array}{rcl}
    Y_{a}(B_i) & \sim & \ell_a(\mu_{a}(B_i), \boldsymbol\theta_{1a}),\\
    g_a(\mu_{a}(B_i)) & = & \sum_{s_k\in B_i}\frac{\mathbf{A}(s_k)\boldsymbol\beta}{n(s_k\in B_i)} + \sum_{s_k\in B_i} \frac{\mathbf{A}_u(s_k) \mathbf{u}(s_k|\rho,\sigma)}{n(s_k\in B_i)},\\
    Y_{b}(C_j) & \sim & \ell_b(\mu_{b}(C_j), \boldsymbol\theta_{1b}),\\
    g_b(\mu_{b}(C_j)) & = & \alpha\cdot\sum_{s_k\in C_j} \frac{\eta_{j,a}(s)}{|C_j|} + u_j, \\ 
    & = &\alpha\left[\sum_{s_k\in C_j}\frac{\mathbf{A}(s_k)\boldsymbol\beta}{n(s_k\in C_j)} ds + \sum_{s_k\in C_j} \frac{\mathbf{A}_u(s_k) u(s_k|\rho,\sigma)}{n(s_k\in C_j)} \right] + u_j,\\
\end{array}
\end{equation}
where the first and second datasets are related to $B_i$ and $C_j$ areas respectively, and the continuous Gaussian field allows us to combine both datasets, which can also be applied to covariates in case we lack their information. An additional challenge arises when both sources of information are provided at an areal level but do not match. In such cases, downscaling can be applied to both datasets to facilitate their joint evaluation, or appropriate aggregation methods can be used when feasible. 

In the case of areal-areal change of support, it is also possible to explore alternative processes beyond the definition of a continuous Gaussian field. Specifically, when one information source is available at a coarser spatial scale than another, an alternative downscaling approach can be considered without explicitly defining a continuous Gaussian field. This approach is particularly useful when the areas associated with the coarser data are entirely nested within a set of smaller areas corresponding to the alternative information source, such that $C_j=\bigcup_kB_k$. In this scenario, the predictor associated with each polygon $C_j$ can be expressed as:
\begin{equation}
\eta_j\propto\sum_{k:B_k\subset C_j}\frac{|B_k|}{|C_j|}\eta_k,
\end{equation}
where the predictor is progressively constructed from finer-scale data and aggregated upward. In the limit, this structure corresponds to cases where information is not directly cross-referenced between scales, i.e., where no change of support or misalignment occurs. This component represents the shared predictor, and when coarser data are available at multiple scales, additional calibration mechanisms can be introduced. This approach thus provides an alternative to continuous Gaussian field–based downscaling, while preserving the ability to align externally based models with observational data and to quantify the uncertainty in external data not explained by the observational model.

\subsection{Temporal change of support}

Temporal change of support implies that one information source does not align in its temporal structure with an alternative data source. This discrepancy can be evaluated by determining whether the data are continuously structured or organised into intervals. A similar issue arises in spatial analysis when space is defined in only one dimension, such as in distance sampling models \citep{Yuan_2017_PointProcessDistanceSampling} or metric graphs \citep{Bolin_2024_MetricGraphs}. 

Therefore, the procedures proposed for the two-dimensional spatial case can also be applied in this setting. In a joint model for a continuous process over time, where different observations are available over a period $\mathcal{T}$, and an alternative dataset provides information at intervals $T_i = (t_{i1}, t_{i2})$ that may or may not fully cover the period $\mathcal{T}$, a structure integrating both sources of information can be written as follows: 
\begin{equation}
\begin{array}{rcl}
    Y_{i,a} & \sim & \ell_a(\mu_{i,a},\boldsymbol\theta_{1a}),\\
    g(\mu_{i,a}) & = &\eta_{i,a}=\mathbf{A}(s)\boldsymbol\beta + u(s|\rho,\sigma),\\
    Y_{j,b} & \sim & \ell_b(\mu_{j,b},\boldsymbol\theta_{1b}),\\
    g_b(\mu_{j,ex}) & = &\alpha\cdot\int_{s\in T_j} \frac{\eta_{j,a}(s)}{|T_j|}ds + u_j, \\
    & = &\alpha\left[\sum_{s_k\in T_j}\frac{\mathbf{A}(s_k)\boldsymbol\beta}{n(s_k\in T_j)} + \sum_{s_k\in T_j} \frac{\mathbf{A}_u(s_k)u(s_k|\rho,\sigma)}{n(s_k\in T_j)} \right] + u_j,\\
\end{array}
\label{eq:temporal_change_support_1}
\end{equation}
where $s$ and $s_k$ denote temporal or one-dimensional spatial locations, and $u_j$ is an additional component that accounts for uncertainty in the scale of the aggregated or coarser data information. This equation is a reformulation of Equation~\eqref{eq:spatial_change_support_1} for the one-dimensional case, which is commonly encountered in temporal analyses. Moreover, it does not impose any regularity on the intervals—they can be irregular and need not represent an average over the interval. Instead, by setting the normalization factor to $1$, the cumulative quantity is obtained.

\subsection{Spatio-temporal change of support}

An extension to the spatio-temporal case can be achieved by combining our previous two proposals (spatial and temporal). For spatio-temporal change of support, we propose procedures analogous to those used for spatial and temporal cases. The key difference is that, in the spatio-temporal case, integration occurs in three dimensions, requiring the approximation of a volume integral. In this setting, the response variable is continuous in both space and time, and an external source provides information within specific volumes of the spatio-temporal domain. Consequently, a possible model that combines an observational dataset $Y_{i,a}$ and external data $Y_{j,b}$ given a spatio-temporal aggregation into voxels ($V_j$) of the external information, can be written as follows:
\begin{equation}
\begin{array}{rcl}
    Y_{i,a} & \sim & \ell_a(\mu_{i,a},\boldsymbol\theta_{1a}), \\
    g_a(\mu_{i,a}) & = & \eta_{i,a}=\mathbf{A}(s)\boldsymbol\beta + u(s), \\
    Y_{j,b} & \sim & \ell_b(\mu_{j,b},\boldsymbol\theta_{1b}), \\
    g_b(\mu_{j,b}) & = & \alpha\cdot\iiint_{s\in V_j} \frac{\eta_{j,a}(s)}{|V_j|}ds + u_j, \\
    & = &\alpha\left[\sum_{s_k\in V_j}\frac{\mathbf{A}(s_k)\boldsymbol\beta}{n(s_k\in V_j)} + \sum_{s_k\in V_j} \frac{u(s_k)}{n(s_k\in V_j)} \right] + u_j, \\
\end{array}
\label{eq:spatiotemporal_change_support}
\end{equation}
where $s$ and $s_k$ represent spatio-temporal locations and $u_j$ is an additional component accounting for uncertainty in the external dataset. This equation represents the extension of the previous spatial and temporal change of support formulations, now accounting for the three-dimensional nature of the data.

\subsection{Categorical change of support}

The latter situation arises when dealing with a categorical change of support. In this case, categorical change of support occurs when an information source is provided using a smaller set of categories derived from the aggregation of a larger set in an alternative data model; the reverse scenario is also possible. This means that category effects can be combined if there is a clear correspondence between groups. 

For instance, consider a categorical variable with $5$ categories, $\mathbf{u} = {u_1, \dots, u_5}$, and an alternative source consisting of data with $3$ categories, $\mathbf{u}^* = \{u^*_1, \dots, u^*_3\}$, such that $u^*_2 = \{u_2, u_3\}$ and $u^*_3 = \{u_4, u_5\}$. In this case, the aggregation in the joint model assumes that in the second dataset, the estimates for $u^*_2$ and $u^*_3$ are equivalent to the sums of the effects of the corresponding categories $u^*_2 = u_2 + u_3$ and $u^*_3 = u_4 + u_5$. In general, this can be stated as follows: if we have a categorical variable with $K$ categories and an external source that provides information for the same variable in $K^*$ categories, where $K > K^*$, and if there is a correspondence between disjoint groups $g_k = \{\mathbf{u}_k\}$ of levels of the variable in the first dataset with respect to the second information source---such that $g_k \in \{u_1, \dots, u_K\}$ and $g_i \cap g_j = \emptyset \iff i \neq j$---then the aggregation can be written as:
\begin{equation}
u^*_k = \sum_{u_i \in g_k} u_i. 
\label{eq:base_balanced_categorical_change_support}
\end{equation}
This approach relies on the underlying assumption that, across the grouped factor levels in the higher structure---i.e., at the lower-resolution scale---the grouping from the higher-resolution scale is balanced.

However, a more general and flexible formulation accounts for unbalanced aggregation. In this case, Equation (\ref{eq:base_balanced_categorical_change_support}) is modified as:
\begin{equation}
u^*_k = \sum_{u_i \in g_k} w_{ik} u_i. 
\label{eq:base_unbalanced_categorical_change_support}
\end{equation}
where $w_i$ represents the weight associated with each factor level $u_i$ in the aggregation $u_i \in g_k$, reflecting its actual proportion in the grouping, given by $n(u_i \in g_k)/\sum_{i=1}^K n(u_i \in g_k)$.  The weights satisfy the constraints $w_{ik}\geq 0$ and $\sum_{i=1}^K w_{ik}=1$.

Accordingly, its implementation, following the model structures previously used, can be expressed as:
\begin{equation}
\begin{array}{rcl}
    Y_{i,a} & \sim & \ell_a(\mu_{i,a},\boldsymbol\theta_{1a}), \\
    g(\mu_{i,a}) & = & \eta_{i,a} = \mathbf{A}_a\boldsymbol\beta + u(z_{i}), \\
    Y_{j,b} & \sim & \ell_b(\mu_{j,b},\boldsymbol\theta_{1b}), \\
    g_e(\mu_{j,b}) & = & \alpha\cdot (\mathbf{A}_{b}\boldsymbol\beta + u^*(g_j)) + u_j, \\
    & = & \alpha\cdot (\mathbf{A}_{b}\boldsymbol\beta + \sum_{z_i \in g_j}u(z_{i})) + u_j, \\
\end{array}
\label{eq:categorical_change_support}
\end{equation}
where $\{\mathbf{A}_a, \mathbf{A}_b\}$ are the design matrices for the first and second information sources, respectively; $u^*(g_j)=\sum_{z_i \in g_k}u(z_{i})$ represents the change of support across different category levels following Equation~\eqref{eq:base_balanced_categorical_change_support}; and $u_j$ is an additional component accounting for uncertainty in the second information source.

A limiting case of this aggregation arises when it is applied to the entire categorical variable. In this situation, the effects of categories not included in the second source should not be shared between the models for the first and second datasets. This is because categorical variables typically satisfy a sum-to-zero constraint, $\sum_{k=1}^K u_k = 0$, where $K$ is the total number of categories. Consequently, the aggregated effect in the second model is also equal to zero. Therefore, only the aggregated information across category levels is retained, and since $\sum_k u_k = 0$, the integration over the entire categorical variable results in zero in the second model.

\section{Examples}

In this section, three examples are presented to illustrate the integration of different information sources under change of support and sequential inference using INLA. The first two examples focus on strategies for handling change of support in different contexts: the first addresses spatial change of support, while the second deals with categorical change of support. The third example demonstrates a case of sequential inference with a spatio-temporal structure, comparing the results obtained from the sequential procedure with those from fitting the model to all data simultaneously.

\subsection{Data integration with spatial change of support}

In this first example, we consider a case in which limited observational data are combined with external information. This simulated example is designed to replicate a structure for incorporating expert knowledge similar to that presented in \citep{Pearman_2020_SpatialExpertElicitation}, where the external data provide information on the probability of presence within survey areas. It also represents situations in which information obtained through expert elicitation reflects the experts’ understanding of the species’ spatial distribution. Such knowledge can be integrated using strategies of this kind \citep{Yamada_ElicitingWildlifeHabitat_2003, Gobbi_ExpertBasedConservation_2012, Aizpurua_HabitSuitabilityExpertKnowledge_2015}.

\begin{figure}[h!]
\includegraphics[width=\linewidth]{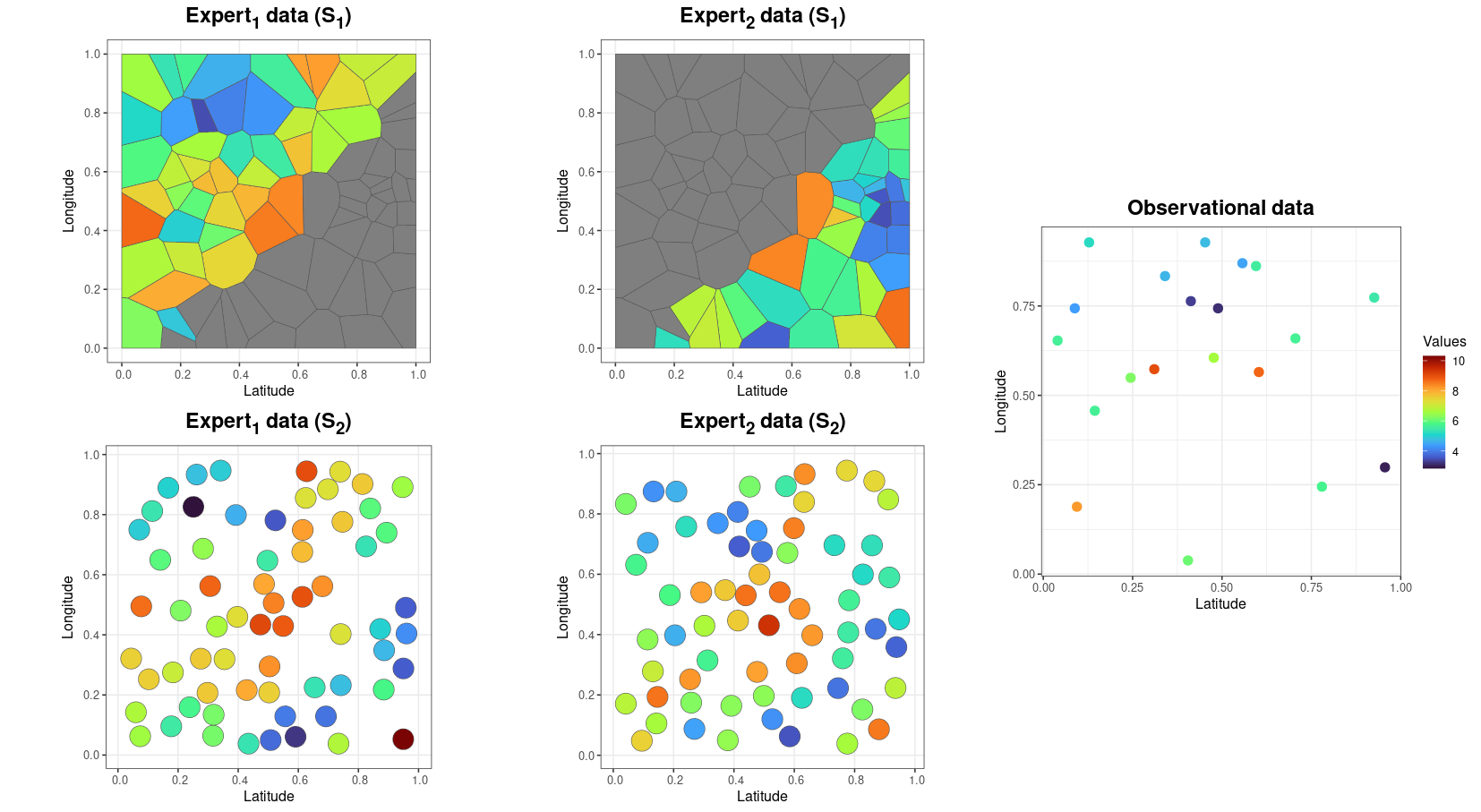}
\caption{Observational data from a limited source and information from external sources according to two different spatial structures.}
\label{fig:observations_expert_data}
\end{figure}

\begin{figure}[h!]
\includegraphics[width=\linewidth]{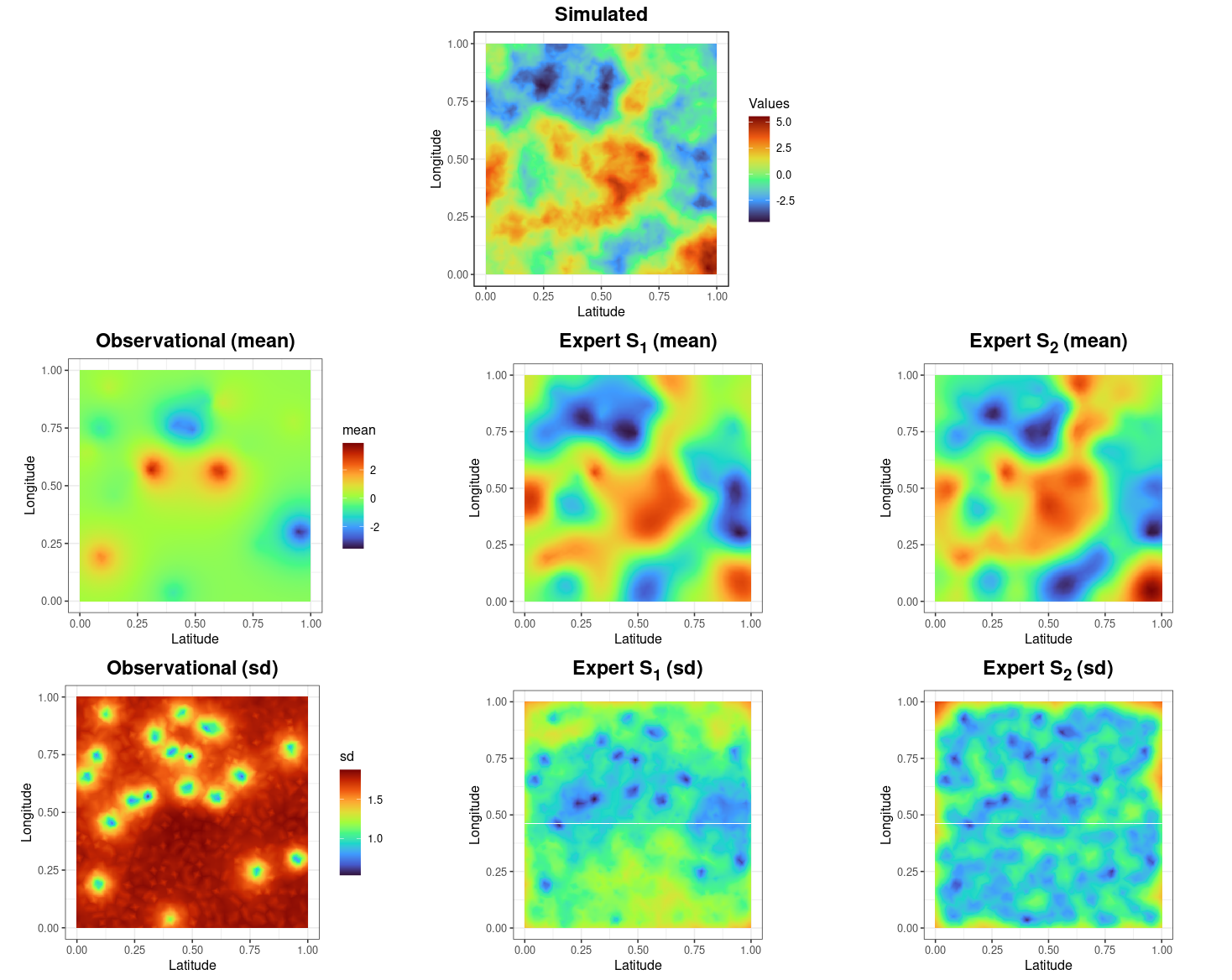}
\caption{Simulated spatial effect, along with the mean and standard deviation of the posterior distribution of the spatial effect, analyzed either using only the observational data or by integrating expert information according to the two structures ($\text{S}_1$ or $\text{S}_2$) for expert information.}
\label{fig:spatial_observations_expert}
\end{figure}

In this specific case, we assume that two experts provide information following distinct spatial structures. This may occur either by defining a predetermined structure that covers the entire study area ($\text{S}_1$), as in \citep{Kaurila_SDMExperElicitationBayesCalibration_2022}, or by allowing each expert to delineate specific areas ($\text{S}_2$) where they can provide information. Figure~\ref{fig:observations_expert_data} illustrates these two spatial structures ($\text{S}_1$ and $\text{S}_2$), together with the available observations in a scenario with limited spatial data. Accordingly, the model used to simulate and infer the data incorporates the spatial change of support between the expert and observational information, as well as the correlation between experts:
\begin{equation}
\begin{array}{rcl}
y_{i,obs} & \sim & \mathcal{N}(\mu_i,\tau),\\
\mu_i & = & \beta_0 + u_s(s_i),\\
\mathbf{y}_{j,ex} & \sim & \mathcal{MVN}(\boldsymbol\mu_j, \boldsymbol\Sigma), \\
\boldsymbol\mu_j & = & \boldsymbol\beta_{0,ex} + \int_{C_j}\frac{\beta_0 + u_s(s)}{|C_j|}\text{d}s,
\end{array}
\end{equation}
where we share the entire latent field between the observational model and the auxiliary expert model. The matrix $\boldsymbol\Sigma$ represents the variance-covariance between the experts, capturing both the marginal uncertainty of the experts and the correlation between them. In Fig. \ref{fig:spatial_observations_expert}, the results of the spatial field are presented, either modeled using only the observational data or integrating expert information using the aforementioned model, according to the two spatial structures in which the experts provided their information.

\subsection{Data integration with categorical change of support}

\begin{figure}
\includegraphics[width=\linewidth]{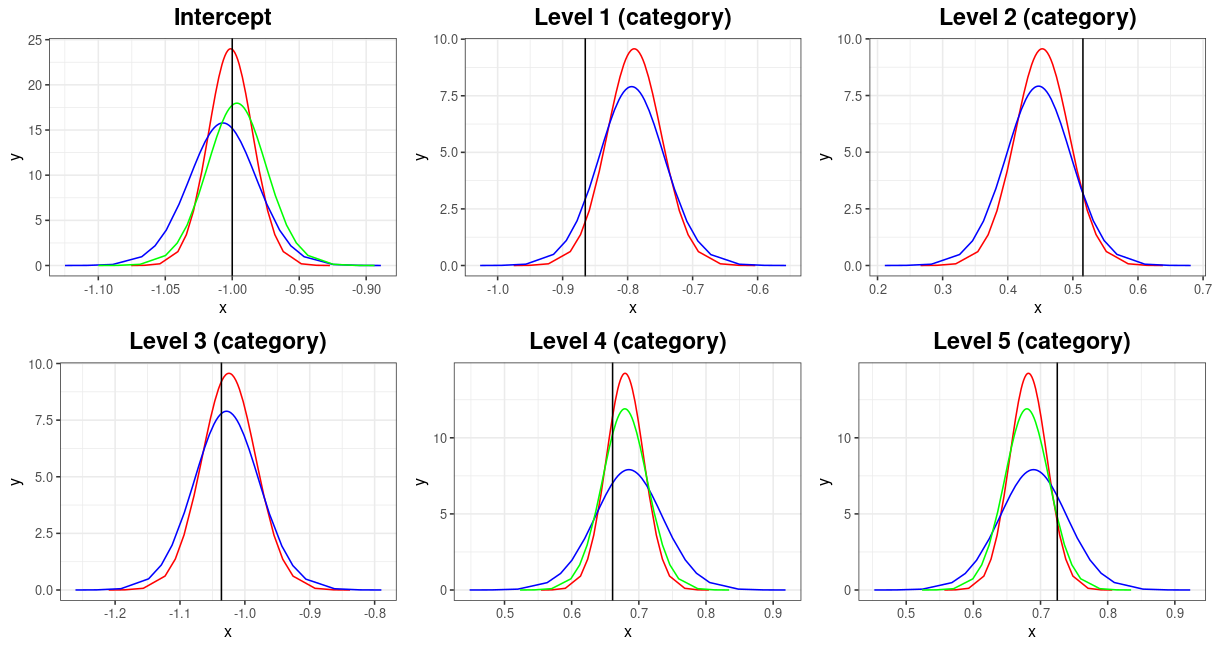}
\caption{Posterior distributions for the intercept and the levels of the categorical variable. The vertical black line indicates the true value, the red line corresponds to the model that integrates both data sources, the blue line is associated exclusively with the first data, and the green line represents the contribution from the second data exclusively (which is why it is not present for the first three categories).}
\label{fig:categoriacal_change_of_support}
\end{figure}

In this example, we assume the existence of two data sources. The first dataset, $\mathbf{y}_a$, contains a categorical variable $\mathbf{u}a$ with five levels, while an alternative source, $\mathbf{y}_b$, provides information structured in three levels for a categorical variable $\mathbf{u}_b$. In this setting, the first three levels of the categorical variable in the first source are aggregated into a single category in the second source, while the remaining two levels are retained unchanged ($u_{b2} = u_{a4}$ and $u_{b3} = u_{a5}$). Thus, the categorical variable in the first model is defined as $\mathbf{u}_{a} = \{u_{a1}, u_{a2}, u_{a3}, u_{a4}, u_{a5}\}$, and in the second source as $\mathbf{u}{b} = \{u_{b1}, u_{b2}, u_{b3}\}$, where $u_{b1}=u_{a1}+u_{a2}+u_{a3}$. Accordingly, the model used is as follows:
\begin{equation}
\begin{array}{rcl}
y_{i,a} & \sim & \mathcal{N}(\mu_i,\tau_{a}),\\
\mu_{i,a} & = & \beta_0 + u_{ai}, \\
y_{j,b} & \sim & \mathcal{N}(\mu_j,\tau_{b}),\\
\mu_{j,b} & = & \beta_0 + u_{bi},
\end{array}
\end{equation}
where information about the intercept and the effects of the categorical levels is shared, considering the aggregation described above. Figure~\ref{fig:categoriacal_change_of_support} presents the results for the intercept and the posterior distributions of each category level.

\subsection{Recursive inference using real spatio-temporal temperature data}

{ In this third example, we illustrate the use of the recursive inference algorithm to perform spatio-temporal modeling of temperature data. The dataset consists of observations from $308$ spatial locations over $480$ temporal nodes, with one node corresponding to each month. The values represent the average monthly temperature (in \textdegree C) for each spatial location along the coast of Alicante, Spain. This example was used in \cite{Figueira_2025_SequentialConsensusInference} to evaluate the performance of the sequential consensus approach where it was shown the benefits of using sequential algorithms for \textit{Big data} contexts. Figure~\ref{fig:Temperature_36_Months} shows the temperature values at the spatial locations for the first 36 months.  

The spatio-temporal model used to analyze the data is as follows:
$$
\begin{array}{c}
y_i \sim \text{N}(\mu_i, \tau), \\
\mu_i = \beta_0 + \mathbf{A}_i \mathbf{u}_{st},
\end{array}
$$
where $\beta_0$ is the intercept, $\mathbf{u}_{st}$ is the vector of latent field parameters associated with the spatio-temporal effect, $\mathbf{A}$ is projection matrix that maps the values from the latent field to the observations, and $\tau$ is the marginal precision of the likelihood. The spatio-temporal effect is defined as $\mathbf{u}_{st} \sim \text{GMRF}(\mathbf{0}, \mathbf{Q}_{st})$, with a precision matrix separable under a Kronecker product structure, $\mathbf{Q}_{st} = \mathbf{Q}_{t} \otimes \mathbf{Q}_{s}$; the first precision matrix ($\mathbf{Q}_{t}$) relates to an autoregressive structure of first order and the second to the matrix ($\mathbf{Q}_{s}$) derived from the SPDE approach. The hyperparameters associated with the spatio-temporal effect are $\rho_s$, $\rho_t$, and $\tau_{st}$, where the first two correspond to the spatial correlation and temporal autocorrelation, respectively, and the last one represents the marginal precision of the spatio-temporal effect. Non-separable spatio-temporal models \citep{Lindgren_2024_NonseparableSpt} can also be implemented within the proposed recursive approach.  

In this example, two cases of data analysis are presented. In the first case, the dataset comprising the first 120 months is analyzed both by modeling all the data jointly and by applying the recursive approach with temporal partitions, grouping the data every 20 months (6 temporal partitions). In the second case, a complete analysis of the entire 480-month dataset is performed using the recursive approach, following the same partitioning criterion, that is, by grouping the data in 20-month intervals.

The first case aims to compare the results and computational performance of the standard method, which fits all the data simultaneously, with those obtained using the recursive method proposed in this work. The second case, in turn, demonstrates that a full joint analysis of all the data is feasible with the recursive approach, whereas it is not possible with the standard method due to memory overflow issues. All analyses were performed on a server with 63 cores and 157 GB of RAM.

In the first case, which compares both approaches, Figure~\ref{fig:marg_hyperpar} shows the marginal posterior distributions of the hyperparameters for the standard full-data analysis (red) and for the recursive approach (blue), where only tiny differences appear between them. Figure~\ref{fig:intercept_random_marginal_spt} presents the posterior marginal distribution for the intercept and for a set of 47 randomly selected nodes of the spatio-temporal effect; again, the marginal from the standard full-data analysis is shown in red, and that from the recursive approach in blue. The differences in the marginal posteriors of the latent field are negligible. Overall, the differences between the recursive and the standard full-data analyses are tiny or negligible. The computation times were $42.87$ minutes for the standard full-data analysis and $10.90$ minutes for the recursive approach, making the latter almost four times faster while producing results that are nearly identical.

In the second case, where it was not possible to run the analysis with the complete dataset of $480$ temporal nodes using the standard full-data method due to memory limitations, the recursive approach successfully produced the results in $48.21$ minutes. These results demonstrate the ability of the recursive approach to achieve outcomes that are almost identical to those of the standard full-data analysis, while providing substantial reductions in computational cost in terms of both time and memory.

\begin{figure}
\includegraphics[width=\linewidth]{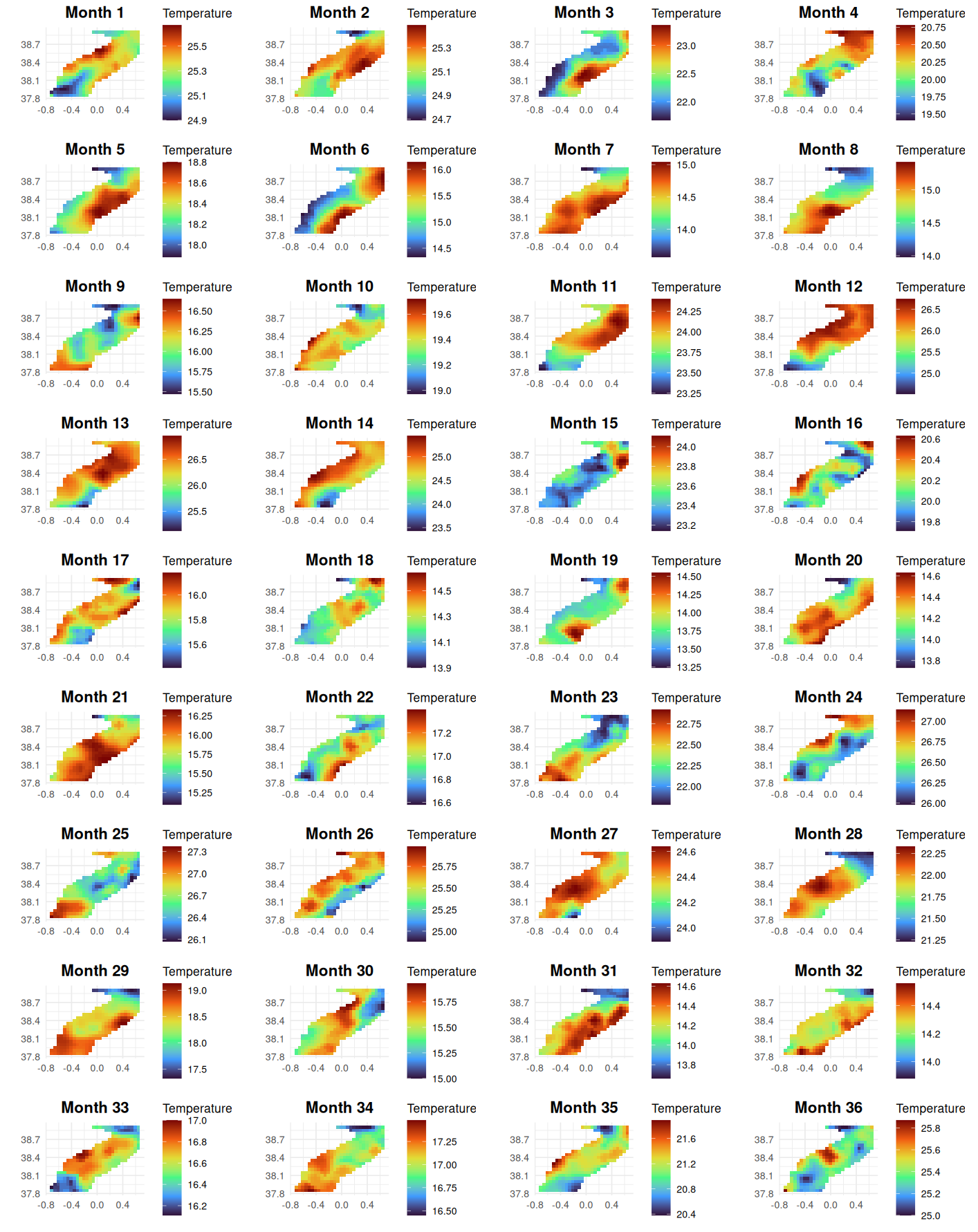}
\caption{ Subsample comprising the first $36$ months (of a total of $480$) from the spatio-temporal temperature (in \textdegree C) dataset.}
\label{fig:Temperature_36_Months}
\end{figure}

\begin{figure}
    \centering
    \includegraphics[width=\linewidth]{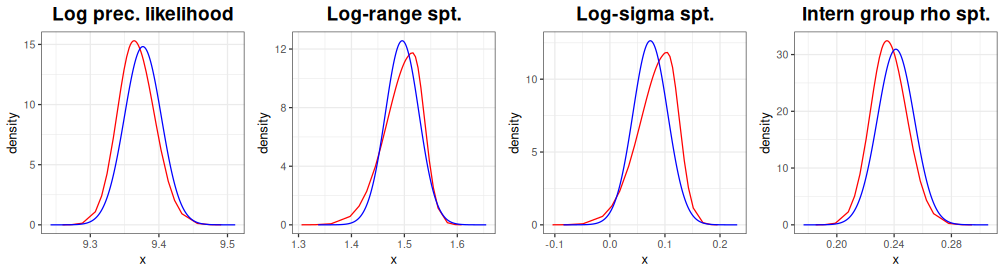}
    \caption{Marginal posterior of the hyperparameters for the standard full-data analysis (red) and from the recursive approach (blue).}
    \label{fig:marg_hyperpar}
\end{figure}

\begin{figure}
    \centering
    \includegraphics[width = \linewidth]{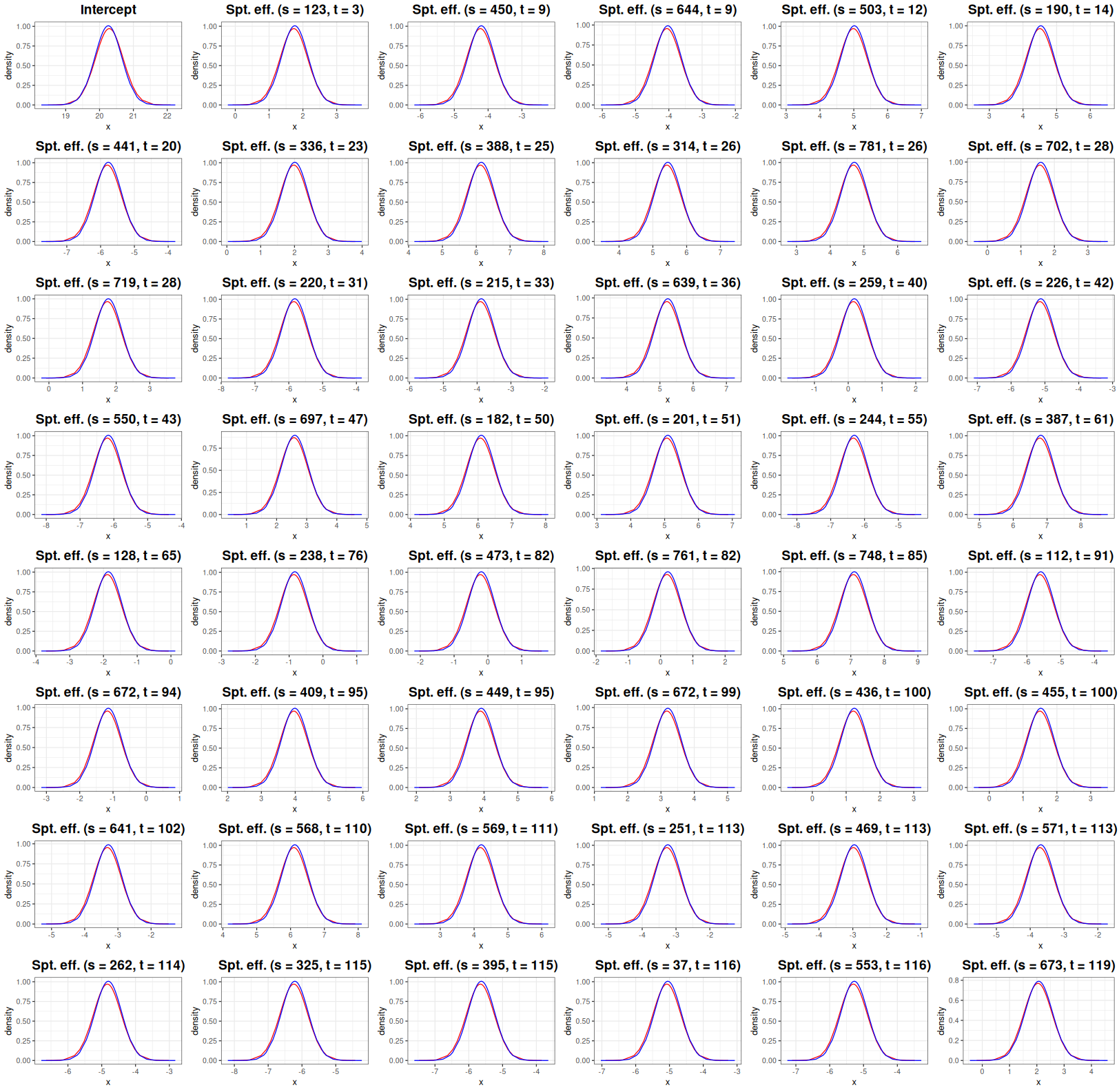}
    \caption{Marginal posterior distributions for the intercept and for 47 randomly selected nodes of the spatio-temporal effect, comparing the standard full-data analysis (red) with the recursive approach (blue). In the labels of the spatio-temporal marginals, $s$ denotes the mesh-node index and $t$ the corresponding temporal index.}
    \label{fig:intercept_random_marginal_spt}
\end{figure}

Finally, it is worth noting that results obtained by this new recursive method solves all the issues detected in the sequential consensus approach as discussed in \cite{Figueira_2025_SequentialConsensusInference}.
}

\section{Conclusions}

{ In this work, we have synthesised several strategies for integrating heterogeneous sources of information and implemented them in \texttt{R-INLA}. We have focused on two main aspects: (i) change of support between different data sources, and (ii) sequential inference strategies for updating models as new data become available. For the change of support problem, we have adopted an integrated modeling approach due to its simplicity. For sequential inference, we have developed a new recursive procedure within the INLA framework, extending prior updating methods and introducing strategies to incorporate information into prior distributions or combine INLA with complementary approaches, while maintaining consistency with its underlying assumptions. Moreover, as these two aspects address distinct challenges, they can be jointly applied.

With respect to the new recursive inferential approach, that leverages the underlying assumptions of INLA, we have highlighted that the determination of support points for the hyperparameters is likely to be slightly different from those determined using the entire dataset. This is due to the Gaussian approximation used to compute the joint marginal posterior distribution of the hyperparameters in the recursive approach. Indeed, by means of different practical examples, we have seen that the resulting posterior distributions of the hyperparameters are nearly the same that those obtained when using the whole dataset at once, but decreasing drastically the computational cost in terms of memory usage and computational times. These examples have also shown the improvement in the posterior distribution of the spatial structure in presence of spatial change of support, as well as in the case of categorical change of support.  

}



\begin{acks}[Acknowledgments]
This paper is part of the results from the HORIZON-CL6-2021-GOVERNANCE-01 project {\sl Land Use and Management modelling for Sustainable Governance} with reference number 101060423. MF, DC and ALQ also thank support by Grant PID2022-136455NB-I00, funded by Ministerio de Ciencia, Innovación y Universidades of Spain and the European Regional Development Fund (MCIN/AEI/10.13039/501100011033/FEDER, UE). Finally, DC acknowledges Grant CIAICO/2022/165 funded by Generalitat Valenciana and Grant RED2024-153680-T also funded by Ministerio de Ciencia, Innovación y Universidades of Spain.
\end{acks}

\bibliographystyle{ba}
\bibliography{bibliography}

\begin{supplement}
The supplementary material presents additional results for the recursive example, as well as an extra example that demonstrates the application of the recursive procedure in a simple spatio-temporal setting. 

The additional results include a comparison of the Central Composite Design (CCD)---the integration scheme for the marginal of the latent field---constructed by the \texttt{R-INLA} software in the standard full-data analysis and the CCD obtained using the recursive inference approach. The supplementary material also provides plots of the spatial posterior mean and standard deviation of the spatial-temporal effects for the first four temporal nodes (months).

The additional example consists of a simulated spatio-temporal scenario. The first part illustrates a recursive analysis performed by temporally partitioning the model and the data, while the second part presents a case in which both the model and the data are partitioned spatially.

Finally, the code to replicate the examples presented in this manuscript is available at \url{https://github.com/MarioFigueiraP/Expert_Elicitation_Recursive_Inference}.
\end{supplement}

\end{document}